\documentclass[prb,twocolumn,aps]{revtex4}
\usepackage[latin1]{inputenc}
\usepackage{latexsym,amstext,amssymb,amsmath}
\usepackage[dvips]{graphicx}
\usepackage{color}
\begin{document}
%%%%%%%%%%%%%%%%

%%%%%%%%%%%%%%%%%%%%%%%%%%%%%%%%%%%%%%%%%%%%%%%%%%%%%%%%%%%%%%%%%
\title{Impact of strain-induced electronic topological transition on
  the thermoelectric properties of PtCoO$_2$ and
  PdCoO$_2$}
%%%%%%%%%%%%%%%%%%%%%%%%%%%%%%%%%%%%%%%%%%%%%%%%%%%%%%%%%%%%%%%%%     
 \author{Markus Ernst Gruner$^{1,2}$}
 \email{Markus.Gruner@uni-due.de}
 \author{Ulrich Eckern$^{3}$}
 \author{Rossitza Pentcheva$^{2}$}
 \affiliation{$^{1}$Forschungs-Neutronenquelle Heinz Maier-Leibnitz (FRM II), Technische Universit\"at M\"unchen, 85748 Garching, Germany}
 \affiliation{$^{2}$Faculty of Physics and Center for Nanointegration, CENIDE, University of Duisburg-Essen, 47048 Duisburg, Germany}
 \affiliation{$^{3}$Institute of Physics, University of Augsburg, 86135 Augsburg, Germany}
 
\date{\today}
%%%%%%%%%%%%%%%%
\begin{abstract}
  By a combination of first-principles calculations %in the framework of density functional theory
  and semi-classical Boltzmann transport theory,
  we investigate the effect of epitaxial strain on the electronic structure and transport properties
  of PtCoO$_2$  and PdCoO$_2$.
  In contrast to the rather uniform elastic response of both systems,
  we predict for PtCoO$_2$  a high sensitivity of the out-of-plane transport properties to strain, which is not present in
  PdCoO$_2$.
  At ambient temperature, we identify a considerable absolute change in the thermopower from
  $-107\,\mu$V/K at $-5$\,\%
  compressive strain to $-303\,\mu$V/K at $+5$\,\% tensile strain.
  This remarkable response is related to 
  distinct changes of the Fermi surface, which involve
  the crossing of two additional bands at a moderate compressive in-plane strain.
  Combining our transport results with available experimental data on electrical and lattice thermal conductivity
  we predict a thermoelectric figure of merit of up to $ZT$$\,=\,$$0.25$ at $T$$\,=\,$$600$\,K for strained PtCoO$_2$.
\end{abstract}
%%%%%%%%%%%%%%
\maketitle
%%%%%%%%%%%%%%%%%%%%%%
\section{Introduction}
%%%%%%%%%%%%%%%%%%%%%%
Recently, the quest for new thermoelectric materials beyond
Bi$_2$Te$_3$ has gained considerable momentum which is  owed to the 
increasing need of highly performing materials for energy harvesting and conversion.
The most promising materials classes identified so far are found among the semiconductors, which allow a careful
adjustment of charge carrier type and concentration in order to achieve the optimum balance between thermopower, electrical,
and thermal conductivity.\cite{cn:Wood88,cn:Snyder08}
Together with the idea 
to exploit low-dimensional quantum-well structures to increase the thermoelectric
performance,\cite{cn:Hicks93PRB}
this leads to a series of new high performance materials which are formed of heterostructures
or quantum dot superlattices.\cite{cn:Venkat01,cn:Poudeu06,cn:Ohta07}
Oxides, such as Na$_x$CoO$_2$-type or
Ca$_3$Co$_4$O$_9$,\cite{cn:Terasaki97PRB,cn:Fujita01JJAP,cn:Shikano03APL,cn:Koumoto06MRS,cn:He11}
represent another promising materials class with the advantage of
non-toxicity and abundance of their components in combination with the chemical and
thermal stability rather than record-breaking performance values.

Low-dimensionality is readily realized in
naturally layered structures, which are for instance found in
a specific class of cobaltates, the delafossites.
Among those hexagonal $AB$O$_2$ compounds, PdCoO$_2$ and PtCoO$_2$ take a
special position, since these are -- unlike most of
the oxides, which are semiconducting or insulating -- very good metallic conductors.\cite{cn:Rogers71,cn:Nagarajan01,cn:Marquardt06}
Such a material usually disqualifies for thermoelectric applications,\cite{cn:Wood88} but
it was discovered early that the conductivity of PdCoO$_2$ and PtCoO$_2$ is highly anisotropic.\cite{cn:Rogers71,cn:Tanaka96JPSJ}
Recent theoretical work\cite{cn:Ong10PRB,cn:Ong10PRL} reports
also a qualitative difference in the thermoelectric properties
with respect to transport within the $a$-$b$ plane and along the hexagonal $c$-axis.
The calculations predict significant negative values of about $-100\,\mu$V\,K$^{-1}$ (PdCoO$_2$)
and $-250\,\mu$V\,K$^{-1}$ (PtCoO$_2$) for the out-of-plane component of the
thermopower at room temperature, in contrast to moderate $+5\,\mu$V\,K$^{-1}$ in-plane.\cite{cn:Ong10PRL}
Such a large anisotropy in the thermopower can give rise to
a considerable laser induced voltage, which can be exploited to design photosensors
based on the transverse thermoelectric effect.\cite{cn:Yan11JAP}

Both compounds are stable and have been synthesized as high quality single crystals employing the so-called methathetical
reaction.\cite{cn:Shannon70,cn:Shannon71,cn:Prewitt71,cn:Tanaka96JPSJ,cn:Tanaka97}
In PtCoO$_2$ and PdCoO$_2$, the Co ion is found in the trivalent Co$^{3+}$ state, which shows a $S$$\,=\,$$0$
low spin configuration, while Pd and Pt
are monovalent.\cite{cn:Shannon71,cn:Tanaka98PhysicaB,cn:Higuchi03JJAP,cn:Higuchi04JJAP,cn:Noh09PRB}
Experiment\cite{cn:Tanaka98PhysicaB,cn:Higuchi98,cn:Hasegawa01MaterTrans,cn:Noh09PRB} and
first principles theory\cite{cn:Seshadri98,cn:Eyert08,cn:Kim09PRB,cn:Ong10PRB} show consistently that
the Fermi level of PdCoO$_2$ is populated by Pd states, while oxygen states are scarce and
Co states are essentially absent.
Photoemission spectroscopy confirms the validity of this picture also for PtCoO$_2$.\cite{cn:Higuchi03JJAP}
Thus the high in-plane conductivity arises from the hybridized noble metal 4d and 5s electrons
which contribute states at the Fermi level, whereas the CoO$_6$ octahedra can be thought to form an insulating layer,
inhibiting transport along the perpendicular $z$-axis.
The absence of hybridized Co states at the Fermi level results in a quasi-two-dimensional electronic arrangement.
This is mirrored by a Fermi-surface, which shows nearly no dispersion along $k_z$ and
thus has a quasi two-dimensional shape.\cite{cn:Eyert08,cn:Noh09PRL,cn:Ong10PRB,cn:Ong10PRL}
In consequence, the Fermi velocities, i.\,e., the
gradients of $k$-resolved band structure at $E_{\rm F}$, are restricted to the $x$-$y$ plane,
while the rather flat Fermi surface along $k_{z}$ prevents a significant
contribution to
% REV>
%{\newtext
  the out-of-plane conductivity
%}
% >REV
$\sigma_{\rm zz}$.
The predicted cross-section of the Fermi surface in the shape of a closed hexagon
was validated by Noh {\em et al.}\cite{cn:Noh09PRL} based on
angular-resolved photoemission spectroscopy (ARPES).

In the present work, we consider PdCoO$_2$ and  PtCoO$_2$ as an intrinsically %  naturally
layered system serving as a simple prototype of a heterostructure considered
for future thermoelectric applications. We determine the dependence of the electronic structure on
epitaxial in-plane strain by means of
first-principles calculations. Using this as input, we obtain the transport properties
in the framework of semi-classical Boltzmann transport
theory in the constant relaxation time approximation.
We identify a remarkable dependence of the electronic transport of
PtCoO$_2$ on epitaxial strain, which stands in clear contrast to the rather
uniform strain response of the elastic properties.
We will show that this stunning discrepancy is related to an
{\sl electronic topological transition}, which, in turn,
is beneficial in improving the thermoelectric performance of the material.
It manifests  in a variation of the out-plane thermopower by a factor of three and
the out-of-plane conductivity
by more than one order of magnitude.

After a short survey of the computational details in Sec.\ \ref{sec:details},
we will shortly review the similarities in the
structural behavior of expitaxially strained PtCoO$_2$ and PdCoO$_2$ in Sec.\ \ref{sec:struct}.
The different strain response of the electronic structure is explained in  Sec.\ \ref{sec:elec},
while Sec.\ \ref{sec:conduct} is devoted to the immediate consequences for in-plane
and out-of-plane conductivity.
Thermopower and a prediction for the thermoelectric figure of merit for the strained materials
are presented in  Sec.\ \ref{sec:trans}.

\section{Numerical details}\label{sec:details}
The electronic structure was investigated
within the framework of density functional theory %\cite{cn:Hohenberg64}
(DFT) employing the
Vienna ab initio Simulation package ({\sc VASP}),\cite{cn:VASP1,cn:VASP2}
which uses a plane wave basis set for the description of
the valence electrons in combination with
the projector augmented wave approach (PAW).\cite{cn:Bloechl94a}
In our calculations, we considered explicitly the
$2s^22p^4$ electrons for O, $3d^84s^1$ for Co, $4p^64d^95s^1$ for Pd and $5d^96s^1$ for Pt,
choosing a cutoff energy of $500$\,eV.
The exchange-correlation part of the Hamiltonian was represented
by the generalized gradient approximation (GGA) of Perdew, Burke and Ernzerhof.\cite{cn:Perdew96}

The response of the system to epitaxial strain exerted along the $a$-$b$-plane
was calculated from the hexagonal 12 atom unit cell within the
scalar-relativistic approximation. The hexagonal basis of this cell allows a straightforward
manipulation of the lattice parameters according to the epitaxial constraint.
For a given value of $a$, the $c$ lattice parameter and atomic positions were optimized
for minimum energy modelling the epitaxial constraint.
In order to obtain accurate eigenvalues, as needed for the transport calculations described below,
we transformed the
12 atom unit cell with hexagonal basis to the 4 atom primitive cell with rhombohedral basis
using a mesh of 41$\times$41$\times$41 $k$-points. We
included the spin-orbit term to the Hamiltonian in our self-consistent treatment,
which yields small but noticeable changes to the results.
For additional corroboration and high quality Fermi surfaces,
parts of the calculations were repeated with the full-potential augmented plane wave method Wien2k.\cite{cn:Wien2k}
Thermoelectric properties at finite temperatures 
were obtained in the framework of semiclassical Boltzmann transport theory in the constant relaxation time approximation
under the constraint of a conserved number of carriers.
This step was carried out with the BoltzTraP\cite{cn:BoltzTraP} code based on the the eigenvalues
obtained from our VASP calculations.
In order to test the effect of additional correlation on the electronic structure and
transport properties, we applied the GGA+U scheme in the rotationally invariant formulation of
Dudarev {\em et al.}\cite{cn:Dudarev98} on the Co $d$ states for selected cases, using different values
of $U-J$ up to $5\,$eV.
Further technical details can be found in the supplementary material.\cite{cn:supplement}

%%%%%%%%%%%%%%%%%%%%%%%%%%%%%%%%%%%%%%%%%%%%%%%%%%%%%%%%%%%%%%%%%%%%%
\begin{figure} % Fig. 1
  \begin{center}  
\includegraphics[width=0.75\columnwidth]{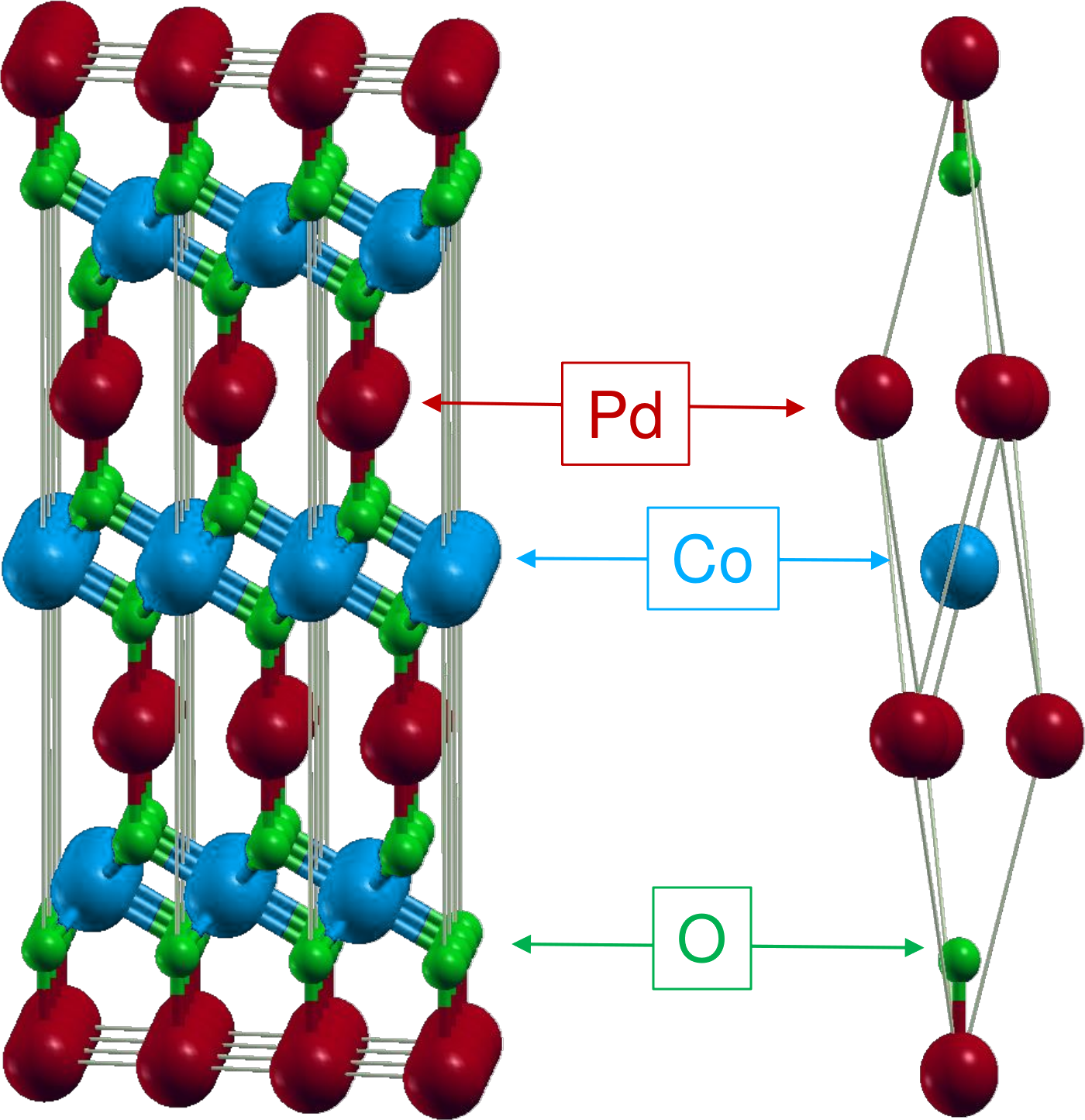}\\
  \end{center}
  \caption{(color online)
    Representation of PdCoO$_2$ and PtCoO$_2$ delafossites in a 
    unit cell with  12 atoms corresponding to the hexagonal basis (left image, 3 times replicated in $a$ and $b$ direction) and a
    four-atom rhombohedral primitive cell (right).
}
\label{fig:cell}
\end{figure}
%%%%%%%%%%%%%%%%%%%%%%%%%%%%%%%%%%%%%%%%%%%%%%%%%%%%%%%%%%%%%%%%%%%%%
\section{Results and discussion}
\subsection{Structural properties strained PdCoO$_2$ and PtCoO$_2$}\label{sec:struct}
The delafossites crystallize in a hexagonal structure %of PtCoO$_2$ and PdCoO$_2$ delafossites
(see Fig.\ \ref{fig:cell}),
which can be described by the rhombohedral space group 166 with symmetry $R\overline{3}m$.
Pt or Pd are found on the (1a) site at $(0,0,0)$, Co on the (1b) site at $(0,0,1/2)$
and the two oxygen on the (2c) site at $(0,0,\pm u)$, with $u$ as an internal structural parameter.
This results in a naturally layered arrangement, consisting of layers of
corner-sharing CoO$_6$ octahedra 
and linear O-Pd-O or, respectively, O-Pt-O dumbbells.
The Pd/Pt atoms form a hexagonal layer with the triangular faces of the CoO$_6$ octahedra
on top of the triangular facets of the Pd/Pt, such that the edge oxygen of CoO$_6$ become part
of the dumbbells.

%%%%%%%%%%%%%%%%%%%%%%%%%%%%%%%%%%%%%%%%%%%%%%%%%%%%%%%%%%%%%%%%%%%%%
\begin{figure} % Fig. 2
  \begin{center}  
\includegraphics[width=0.85\columnwidth]{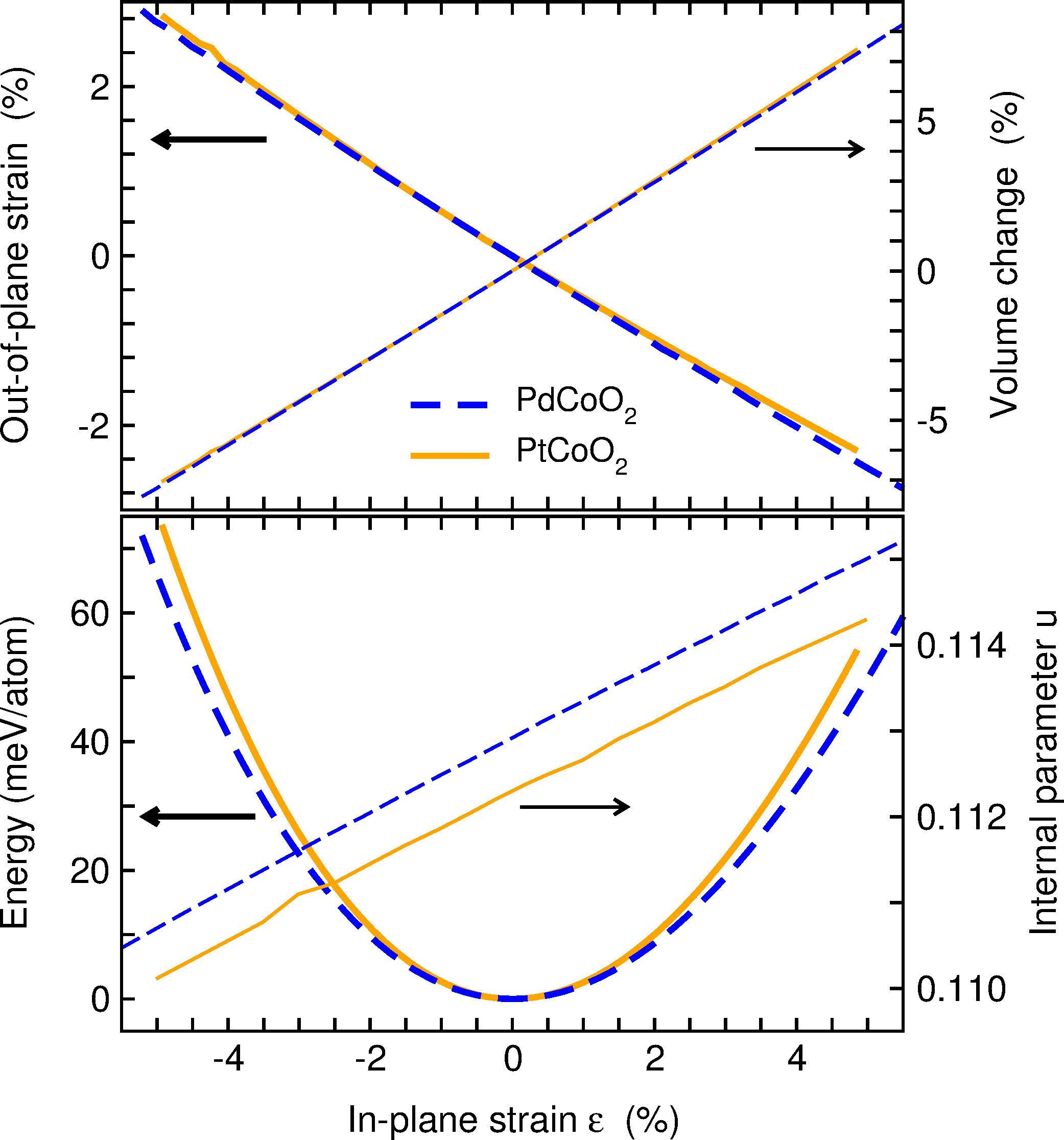}\\
  \end{center}
  \caption{(color online)
    Comparison of total energy (bottom graph, left axis), internal parameter $u$ (bottom graph, right axis),
    out-of-plane strain (top graph, left axis) and unit cell volume (top graph, right axis) relative to the
    respective bulk equilibrium value as a function of the relative in-plane strain for PdCoO$_2$
    (dark blue color) and PtCoO$_2$ (bright orange color)
    obtained from our first-principles calculations.
}
\label{fig:Ecvrel}
\end{figure}
%%%%%%%%%%%%%%%%%%%%%%%%%%%%%%%%%%%%%%%%%%%%%%%%%%%%%%%%%%%%%%%%%%%%%
We find the ground state of both delafossites at nearly the same lattice parameters, i.\,e.,
$a$$\,=\,$$2.870\,$\AA{} and $c$$\,=\,$$17.94\,$\AA{} for PdCoO$_2$ and
$a$$\,=\,$$2.861\,$\AA{} and $c$$\,=\,$$17.95\,$\AA{} for PtCoO$_2$.
Consequently, the atomic volume of the Pd-compound is with
$V$$\,=\,$$10.66\,{\rm\AA}^3$/atom slightly larger than the volume of the Pt-compound
($V$$\,=\,$$10.61\,{\rm\AA}^3$/atom).
These values agree well with previous DFT investigations\cite{cn:Eyert08,cn:Kumar13} and 
experiment ($a=2.83\,$\AA{} for both compounds),\cite{cn:Shannon71,cn:Prewitt71,cn:Tanaka97,cn:Hasegawa03SSC,cn:Takatsu07JPSJ}
with the typical
over-estimation of the lattice constant by approximately one percent as a consequence of the use of the GGA
for the exchange-correlation potential.
Both systems show an essentially identical response to epitaxial strain.
The lower panel of Fig.\ \ref{fig:Ecvrel} indicates that epitaxial growth with in-plane
strains along $a$ of up to $5\,$\% might be realistic, since the corresponding deformation energies
are still in the range of typical thermal energies.
Applying in-plane strain results in a corresponding opposite strain of
the out-of-plane lattice parameter $c$, which is, however, approximately only half as large.
This has the consequence that upon straining the system by $\varepsilon$$\,=\,$$\Delta a/a$$\,=\,$$\pm\,4\,$\%
the equilibrium volume changes significantly by $\Delta V/V$$\,=\,$$\pm\,5\,$\%.
The increase of $c/a$ with decreasing volume is consistent with pressure experiments on PdCoO$_2$.\cite{cn:Hasegawa03SSC}
Epitaxial strain also causes a variation of the internal parameter $u$, which determines the oxygen position
relative to the other ions. At equilibrium conditions, we find $u$$\,=\,$$1.1129$ for PdCoO$_2$ and 
$u$$\,=\,$$1.1123$ for PdCoO$_2$. This corresponds to a Pd-O distance of $2.026$\,\AA{} and a
Pt-O distance of $2.015$\,\AA{}
in combination with Co-O distances of $1.917$\,\AA{} and  $1.919$\,\AA{}, respectively.
Four percent of tensile strain causes $u$ to increase by approximately $2$\,\%.
This has an opposite effect on the Co-O distance as compared to the distance between O and the Pt-group metal.
For both oxides, the Co-O distance increases by $+1.8$\,\%, while the distance between O and the noble metal decreases by
approximately $-0.5$\,\%. Under $-4$\,\% compressive strain, we observe the opposite effect with variations of
$-1.5$\,\% and $+0.6$\,\%, respectively.

\subsection{Electronic structure and Fermi surface of strained PtCoO$_2$ and PdCoO$_2$}\label{sec:elec}
%%%%%%%%%%%%%%%%%%%%%%%%%%%%%%%%%%%%%%%%%%%%%%%%%%%%%%%%%%%%%%%%%%%%%
\begin{figure} % Fig. 3
  \begin{center}  
\includegraphics[width=\columnwidth]{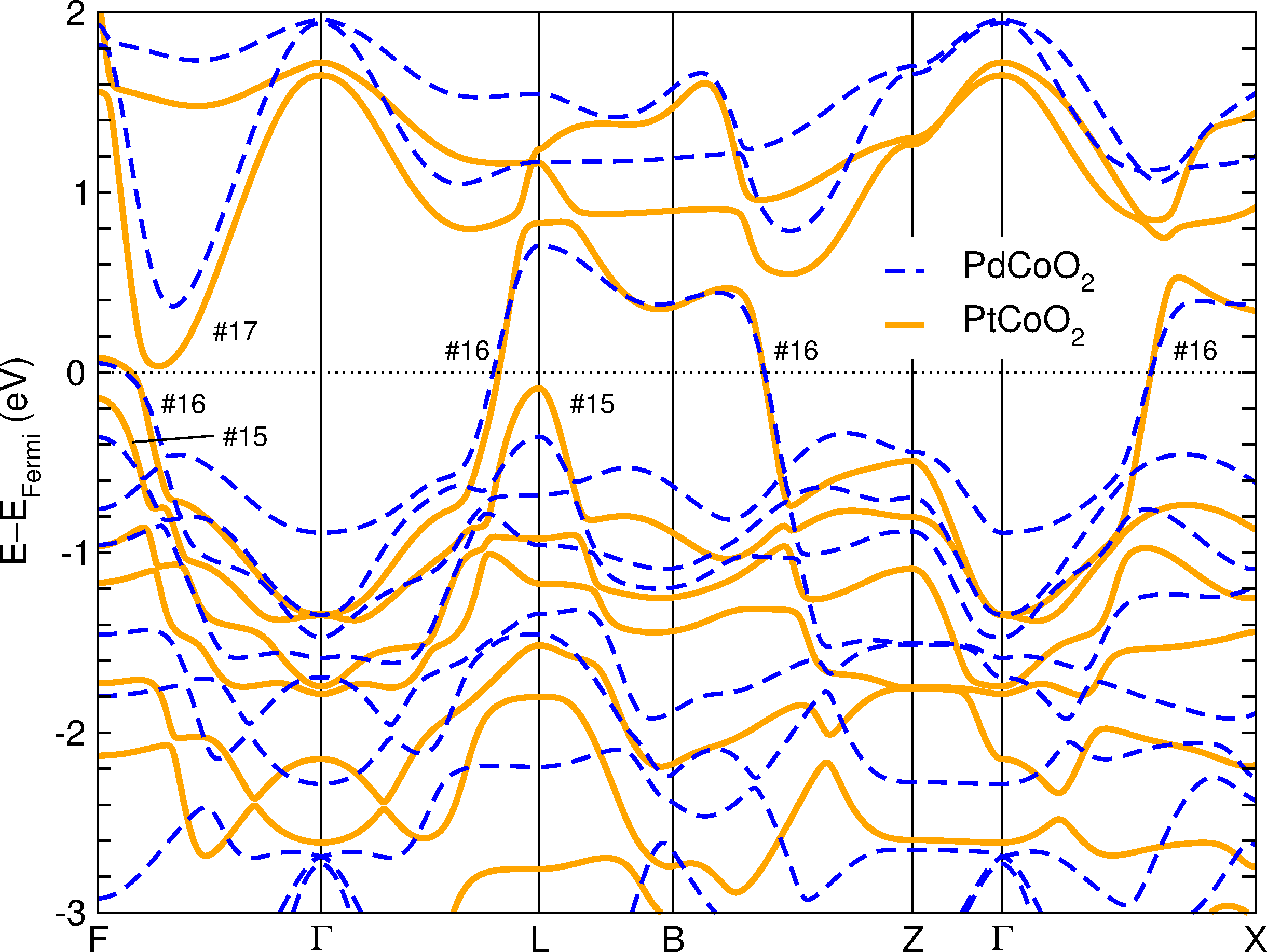}
  \end{center}
  \caption{(color online)
    Band structure of PtCoO$_2$ (solid orange line) and PdCoO$_2$ (dashed blue line)
    in the vicinity of the Fermi level. The numbers denote the bands close to $E_{\rm F}$, which are
    relevant for our discussion. The counting starts from the valence band minimum
    (core and semicore states such as the Pd $4p^6$ are not considered).}
\label{fig:BandBoth}
\end{figure}
%%%%%%%%%%%%%%%%%%%%%%%%%%%%%%%%%%%%%%%%%%%%%%%%%%%%%%%%%%%%%%%%%%%%%
The band structure of the unstrained delafossites, compared in Fig.\ \ref{fig:BandBoth}, suggests
at first sight that the electronic features relevant for the transport properties,
which appear close to the Fermi level, $E_{\rm F}$, are similar for both compounds.
%Fig.\ \ref{fig:BandBoth} shows that
In both cases only one band (i.\,e., the 16th band of valence electrons
disregarding the semicore states in counting) is crossing $E_{\rm F}$
which is formed by $d_{3z^2-r^2}$ and $s$-type orbitals with Pd/Pt character.\cite{cn:Seshadri98,cn:Eyert08,cn:Ong10PRB}
Its rather steep slope in $k_x$-$k_z$-direction, which is responsible for the excellent in-plane conductivity,
corresponds to a section of a parabolic structure,
which has its mimium at $\Gamma$ at $-2.6$\,eV for PtCoO$_2$ and $-2.3$\,eV for PdCoO$_2$
and is intersected by hybridizing bands with d-character several times below $E_{\rm F}$.
Although most other bands of PtCoO$_2$ are shifted downward due to the larger bandwidth
related to the larger extent of the $5d$ shell of Pt compared to the $4d$
electrons of Pd, the 16th band does not change its shape in the immediate vicinity of $E_{\rm F}$.

This is different for the next lower, completely filled band (i.\,e., the 15th band),
which forms hole pockets in the vicinity of the zone boundary,
in particular, near the special points F and L.
The next higher unoccupied band (17th band) displays a marked minimum (electron pocket) along
$\Gamma$-L, which comprises Co $e_{\rm g}$ states\cite{cn:Seshadri98,cn:Eyert08}
that are absent in the vicinity of $E_{\rm F}$ otherwise.
In PtCoO$_2$, these features of the 15th and the 17th band
come very close to $E_{\rm F}$ within $100\,$meV, while in PdCoO$_2$, they
are separated from the Fermi surface by more than $300\,$meV.
As we will show below, these fine details of the electronic structure give rise to
qualitatively different transport properties under epitaxial strain which stand
in contrast to the rather uniform elastic response of both systems.

%%%%%%%%%%%%%%%%%%%%%%%%%%%%%%%%%%%%%%%%%%%%%%%%%%%%%%%%%%%%%%%%%%%%%
\begin{figure*}  % Fig. 4
  \begin{center}  
\includegraphics[width=0.85\textwidth]{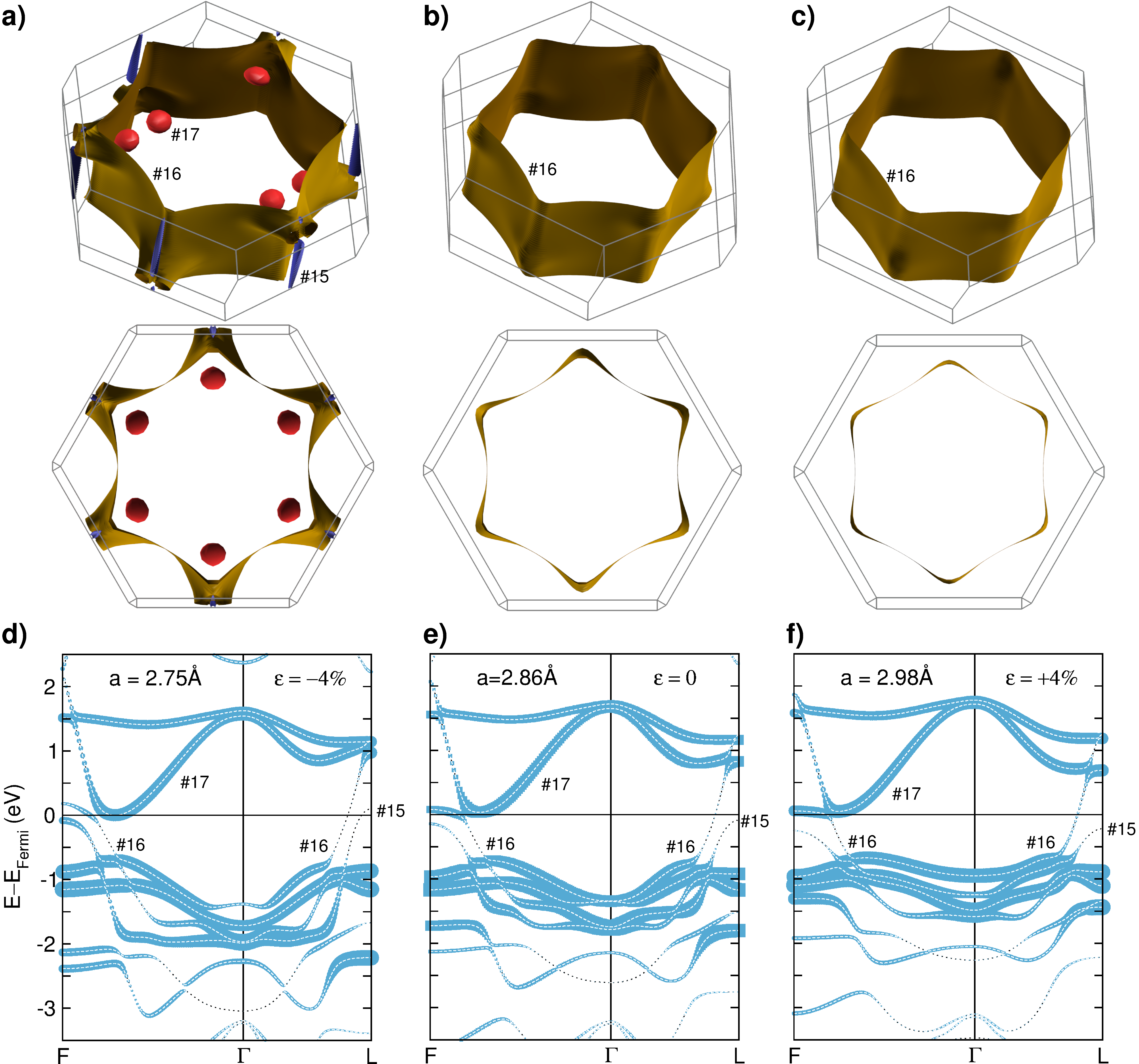}
  \end{center}
  \caption{(color online)
    Fermi surfaces (top and center rows, subfigures a-c) and corresponding band structure (bottom row, subfigures d-f)
    of strained and unstrained PtCoO$_2$.
    The left column (subfigures a and d) refers to $\varepsilon$$\,=\,$$-4$\,\% compressive epitaxial strain, the right column
     (subfigures c and f) to  $\varepsilon$$\,=\,$$+4$\,\% tensile strain and
    the center column  (subfigures b and e) to the unstrained case.
    In the images of the Fermi surfaces the different colors refer to the different bands crossing the Fermi levels
    (blue: 15th band, orange: 16th band, red: 17th band).
    The vertical width of the (bright blue) fat bands in the bottom row
    is proportional to the Co-character of the band. Again, the bands crossing $E_{\rm F}$ are denoted by
    their respective number.
    We find that the 15th and 17th band cross the Fermi energy at compressive strains, whereas for tensile strains,
    the dispersion of
    the 16th band at $E_{\rm F}$ is influenced by the nearby crossing of a parabolic section of the 16th
    Pt-band and the 17th band which is dominated by Co states.
}
\label{fig:FSstrain}
\end{figure*}
%%%%%%%%%%%%%%%%%%%%%%%%%%%%%%%%%%%%%%%%%%%%%%%%%%%%%%%%%%%%%%%%%%%%%

The energy minimum along $\Gamma$-F in the 17th band might be interpreted as a consequence
of the hybridization between the Co $3d_{{\rm e}_{\rm g}}$ bands
with the the parabolic section of the 16th band
(see the Co-resolved band structure of PtCoO$_2$ in Fig.\ \ref{fig:FSstrain}d-\ref{fig:FSstrain}f),
which manifests in a sharp peak in the electronic density of states
$0.4$\,eV above the Fermi level (see supporting material\cite{cn:supplement} and Refs.\ \onlinecite{cn:Seshadri98} and
\onlinecite{cn:Eyert08} for the density of states and orbital resolved band-plots), while the Co
$3d_{{\rm t}_{\rm 2g}}$ and the Pd $4d_{xz}$ and $4d_{yz}$ states are located $0.5$-$2$\,eV below $E_{\rm F}$.
In comparison, the hybridization between Pt $5d_{3z^2-r^2}$ and Co $3d_{{\rm e}_{\rm g}}$ states is less pronounced
and the band-crossing is encountered in the immediate vicinity of the Fermi surface. %cf.\ (Fig.\ \ref{fig:BandBoth})
A strain-induced change of the Brillouin-zone can influence the position of this crossing and
thus show significant impact on the shape of the Fermi surface.
This might be exploited to alter transport properties by applying external fields, such as
mechanical stress.

The crossing of the 16th band results in a hexagonal rod-like shape of the Fermi surface
(Fig.\ \ref{fig:FSstrain}b), which is a pronounced feature of both
compounds.\cite{cn:Eyert08,cn:Noh09PRL,cn:Kim09PRB,cn:Ong10PRB,cn:Ong10PRL}
Under a moderate compressive epitaxial strain of $\varepsilon$$\,=\,$$-4$\,\%,
the extremal features of the 15th and 17th band of PtCoO$_2$ touch the Fermi surface
and changes its topology significantly. This is depicted in Fig.\ \ref{fig:FSstrain}a and Fig.\ \ref{fig:FSstrain}d.
The (blue) needle-type features outside the (orange) hexagonal rod evolve from the rising of the
hole pocket in the 15th band, which is best seen at the L-point. At the same time,
the 17th band approaches the Fermi level along $\Gamma$-F from above resulting in
the (red) pills inside the hexagonal rods.
According to its Pt-type band-character (not shown), we expect
the 15th band to contribute predominantly to the in-plane conductivity.
In turn, the 17th band shows a strong Co-character close to its minimum where the electron pocket forms
and is consequently expected to increase the out-of-plane conductivity (see Sec.\ \ref{sec:conduct}).
While the faces  of the hexagon formed by the 16th band
are essentially dispersionless irrespective of the strain, we find
bulb-like extrusions at its rounded corners.
The extrusions are also present on the Fermi surface of PdCoO$_2$ and
are again associated to a slight admixture of Co-states in band 16 along $\Gamma$-F.\cite{cn:Kim14JPSJ}
In PtCoO$_2$, these features grow under compressive strain and eventually touch
the Brillouin zone boundary, while
they become significantly flatter under tensile strain, in turn. This corresponds to
a diminished band velocity component in $z$-direction.
Tensile epitaxial strain is thus a suitable possibility to control the
Pd/Pt-character of the 16th band and thus enhance the twodimensional nature
of the electronic structure.
In reverse, compressive strain can stimulate an electronic
topological transition, which arises from the the appearance of additional bands at the Fermi level
and the 16th band touching the Brillouin zone boundary.
In PdCoO$_2$, the distance between the Co- and Pd-states along $\Gamma$-F is significantly larger,
which makes the electronic states at the Fermi level much less susceptible to strain.
However, electron doping may provide an alternative way to bring the systems close to this instability and
produce a Fermi surface of similar shape.\cite{cn:Felser99JMaterChem}

Based on experimental evidence,
PtCoO$_2$ and PdCoO$_2$ were considered as weakly correlated oxides.\cite{cn:Itoh99PhysicaB}
Furthermore, Ong {\em et al.}\cite{cn:Ong10PRB,cn:Ong10PRL} found a very
good agreement between the first-principles electronic structure and experimental spectroscopy,
which justifies the use of conventional GGA for the exchange and correlation
part of the Hamiltonian.
However, recently, Hicks {\em et al.} argued that static Coulomb repulsion in terms of a Hubbard model with an
effective $U$-parameter on the Co $3d$ states
improves the agreement with their de-Haas-van-Alphen data.\cite{cn:Hicks12PRL} 
Therefore, we carried out additional GGA+$U$ calculations for PdCoO$_2$ and PtCoO$_2$
under a systematic variation of the effective Coulomb repulsion $U_{\rm eff}$$\,=\,$$U-J$ on Co $3d$ from $0$ (pure GGA) to $5$\,eV.
The additional term mainly affects the bonding $3d$ Co ${{\rm t}_{\rm 2g}}$ states below $E_{\rm F}$ which move 
further down, while we observe only a minute shift in the position of the relevant
Co ${{\rm e}_{\rm g}}$ states, which are close to $E_{\rm F}$. This separates occupied and unoccupied Co and O states further,
but hardly affects the Pd-states at the Fermi surface, which are responsible for metallicity.
Thus $U_{\rm eff}$ has only minor effect on the position of the Pt-dominated 15th band
but larger values will eventually inhibit the crossing of the 17th band or shift it to larger strains.

%%% REV>
%{\newtext
  It is worthwhile to note that the occurrence of the
  electronic topological transition is independent of the
  calculation method and likewise reproduced with VASP and Wien2k. Nevertheless, concerning
  the transport calculations we obtain slight differences between the codes. Interestingly, we find a
  closer agreement of the GGA results obtained from the Wien2k calculations with
  the VASP calculations for $U_{\rm eff}$$\,=\,$$1\,$eV, rather than for the pure GGA,
  although the same exchange-correlation functional (PBE) was used in both cases. We ascribe
  this to the specific implementation of the Co PAW-potential employed in the VASP calculations,
  which apparently allows for a somewhat stronger interaction between Co and Pd states
  resulting in a slightly increased
  presence of residual Co-states at the Fermi-level compared to the full-potential calculations.
  For additional details see the supplementary information.\cite{cn:supplement}
%}
%%% <REV

\subsection{Electronic conductivity under epitaxial strain}\label{sec:conduct}
%%%%%%%%%%%%%%%%%%%%%%%%%%%%%%%%%%%%%%%%%%%%%%%%%%%%%%%%%%%%%%%%%%%%%
\begin{figure} %Fig. 5
  \begin{center}  
\includegraphics[width=0.8\columnwidth]{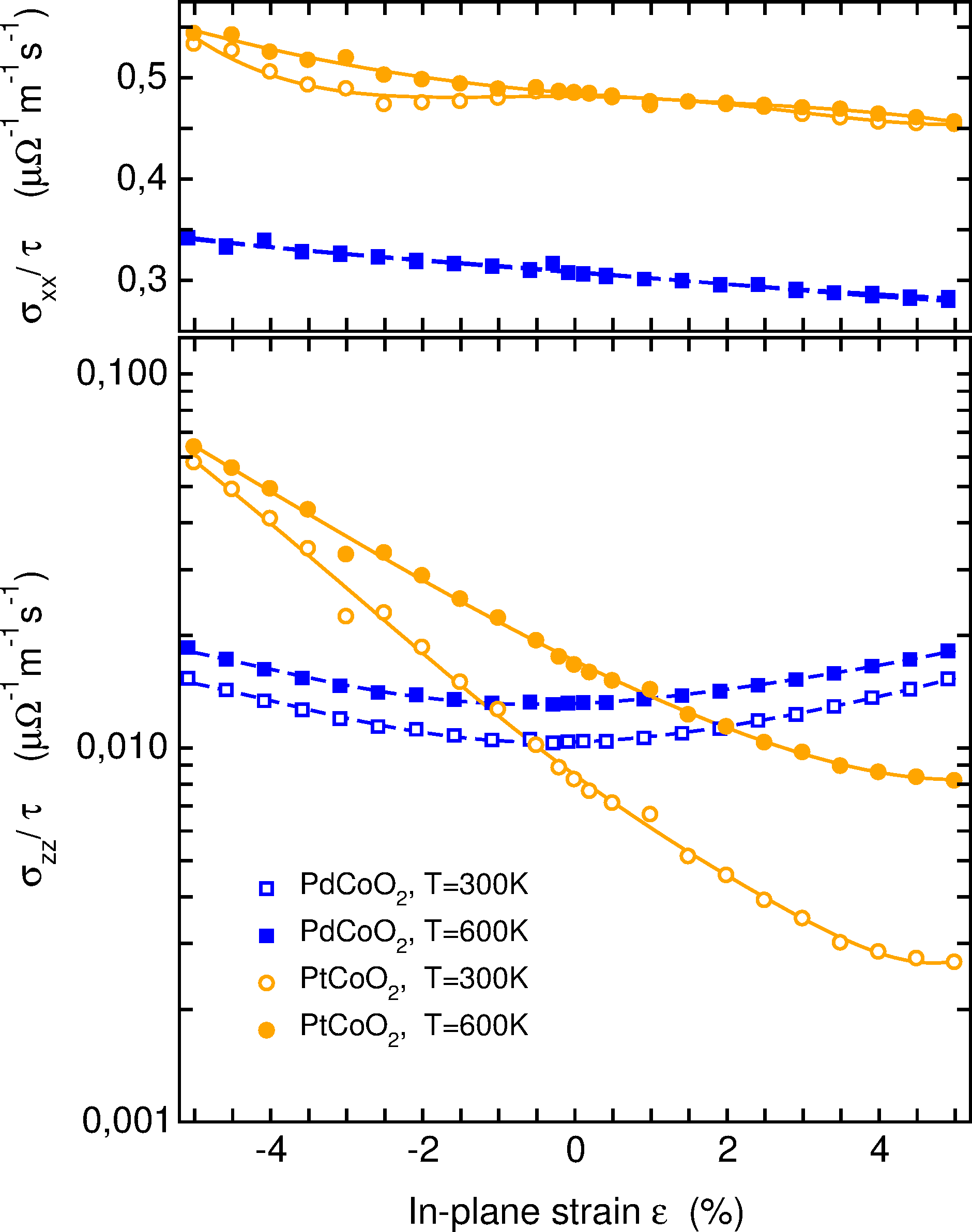}
  \end{center}
  \caption{(color online)
    In-plane and out of plane
    components of the electrical conductivity tensor, $\sigma_{\rm xx}$ (upper panel) and
    $\sigma_{\rm zz}$ (lower panel), as a function
    of epitaxial strain.
    The results are specified relative to the
    empirical parameter of Boltzmann transport theory, the
    relaxation time constant $\tau$.
    We compare the two different compounds
    PdCoO$_2$ (dashed lines and squares) and PtCoO$_2$ (solid lines and circles)
    are compared at two different temperatures $T$$\,=\,$$300\,$K (open symbols) and
    $T$$\,=\,$$600\,$K (filled symbols).    
    Please note the semi-logarithmic scale in the bottom panel.
}
\label{fig:Sigma}
\end{figure}
%%%%%%%%%%%%%%%%%%%%%%%%%%%%%%%%%%%%%%%%%%%%%%%%%%%%%%%%%%%%%%%%%%%%%
Starting from the electronic eigenvalues obtained with DFT, we determined the transport properties 
in the framework of semi-classical Boltzmann transport theory using
the constant relaxation time approximation. This approach
has evolved as a standard tool for the prediction and identification of qualitative trends in a
wide variety of materials,\cite{cn:Scheidemantel03PRB,cn:Bjerg11} including the prediction of
oxidic systems,\cite{cn:Delugas13PRB,cn:Wilson07PRB}
and has been applied successfully to several delafossites in the past.\cite{cn:Singh07PRB,cn:Maignan09PRB}
Concerning PdCoO$_2$ and PtCoO$_2$, particular emphasis was laid on the
explanation of the large anisotropy in conductivity and
thermopower (Seebeck coefficient),\cite{cn:Ong10PRB,cn:Ong10PRL,cn:Kim14JPSJ}
but also on the description of the
large out-of-plane magnetoresistance encountered in PdCoO$_2$ under the rotation of a large
in-plane magnetic field.\cite{cn:Takatsu13PRL}
%%% REV>
The relaxation time approximation, with a single energy- and momentum-independent relaxation time,
assumes that the combined effect of all scattering processes is such that the electronic system
relaxes back to equilibrium exponentially with a single time constant once the perturbation is
switched off. Alternatively, one may consider this approach as a first step towards the complete
description which clarifies the {\em band structure related} aspects of transport. In any case, the
relaxation time approximation allows for solving the Boltzmann equation in a simple way, and hence
an accurate determination of transport coefficients using $10^4$ to $10^5$ $k$-points, which are
necessary for converged results. From the comparison with experiment, on the other hand, one is then
able to deduce information on the relevant scattering mechanisms.

A more detailed theoretical description could be based on Boltzmann theory, but would require a
first-principles calculation of phonon dispersions as well as electron-phonon scattering matrix
elements, which is beyond the scope of the present work. Alternatively, the determination of
transport coefficients could be based directly on a Green's function approach and Kubo's
linear response theory (see Ref.\ \onlinecite{cn:Ebert11} for a recent discussion).

At room temperature and above, it can be expected that electron-phonon scattering is the dominant
relaxation mechanism. For the nearly-free electron model, the scattering operator indeed has been
studied in detail.\cite{cn:Ziman60} For example, it is well known that for temperatures below the
Debye temperature, the momentum and energy relaxation rates are very different, the latter being
much shorter than the former. Above the Debye temperature, electron-phonon scattering is essentially
elastic, and both rates are very similar. These results rely, in particular, strongly on the fact
that the electronic density of states (DOS) is practically independent of energy around the Fermi
level. Thus predictions based on an energy-independent $\tau$ are likely to fail in case sharp
features exist close to $E_{\rm F}$.
This was shown recently for the simple metal Li, where the constant relaxation time
approximation predicts the wrong sign of the thermopower.\cite{cn:Xu14}
For Li, quantitative agreement can be achieved within a variational approach
to Boltzmann theory.\cite{cn:Allen78,cn:Savrasov96,cn:Xu13,cn:Xu14}
Since our calculated transport properties are consistent with
the available experimental data for PdCoO$_2$, as shown below,
we do not expect significant new insights from an advanced treatment of the Boltzmann equation.
However, our results will demonstrate clearly that -- according to the anisotropic nature of the lattice
structure -- one must take into account at least a directional dependence of $\tau$ in
addition to its variation with temperature.
%%% <REV

In practice, all transport tensors
obtained from Boltzmann transport theory
turn out to be essentially diagonal;
there are only two distinct entries. The first, marked with
the index ``xx'', corresponding to in-plane transport, the second, marked as ``zz''
corresponding to transport properties perpendicular to the Co-O and Pd-O/Pt-O layers.

As already anticipated, the significant changes at the Fermi surface of PtCoO$_2$
should leave a corresponding signature in the strain-dependent conductivity, shown in
Fig.\ \ref{fig:Sigma}. 
The variation of the in-plane component, which accounts for
the already very good conductivity in this direction,  is not substantial (upper panel).
According to the steeper slope at $E_{\rm F}$, we find a larger ratio $\sigma_{\rm xx}/\tau$ for PtCoO$_2$ compared
to PdCoO$_2$. Both values are steadily reduced for tensile strains.
However, as indicated earlier, the conductivity is highly anisotropic and thus orders of magnitude
smaller in the perpendicular direction.
According to the changes in the Fermi surface, the strain induced variation of the out-of-plane conductivity
is much more significant for bulk PtCoO$_2$.
Expanding the $a$-axis by 4\% decreases the out-of-plane conductivity by almost one order of magnitude.
In turn, the opposite trend is encountered for compressive strain.
This significantly increases the conductivity compared to PdCoO$_2$, where the variations are much smaller
and do not exhibit a consistent trend for both strain directions.

%%%%%%%%%%%%%%%%%%%%%%%%%%%%%%%%%%%%%%%%%%%%%%%%%%%%%%%%%%%%%%%%%%%%%
\begin{figure} % Fig. 6
  \begin{center}  
\includegraphics[width=0.8\columnwidth]{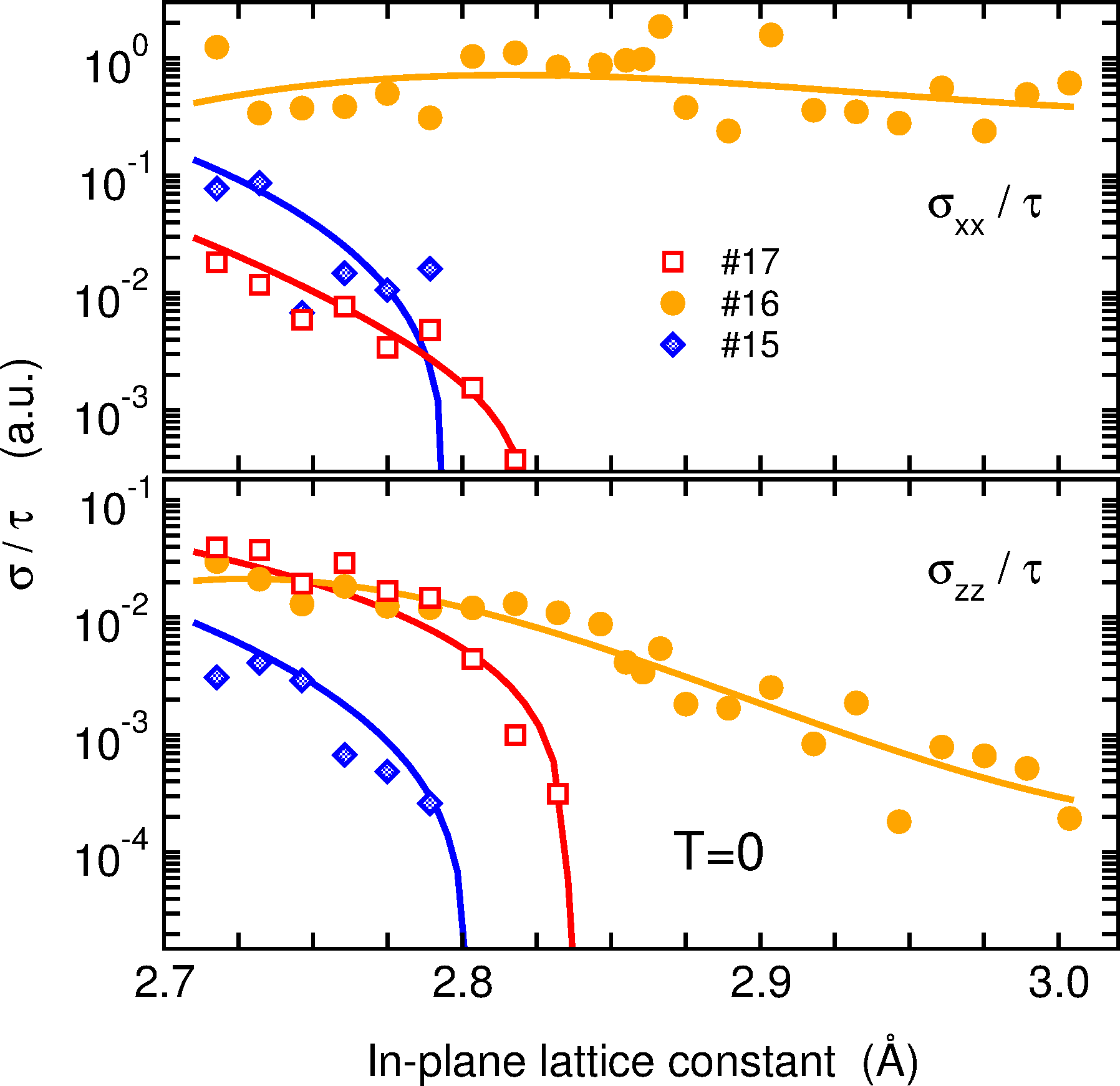} %{fig/Pt-sigtens-a.png}
  \end{center}
  \caption{(color online)
    Band-resolved diagonal components $\sigma_{\rm xx}$ and $\sigma_{\rm zz}$ of the
    electronic conductivity tensor at $T$$\,=\,$$0$ of PtCoO$_2$ as a function of the
    in-plane lattice constant $a$.  The values are specified relative to the
    relaxation time constant $\tau$.
    Only three bands crossing $E_{\rm F}$ contribute to the transport properties.
}
\label{fig:sigbands}
\end{figure}
%%%%%%%%%%%%%%%%%%%%%%%%%%%%%%%%%%%%%%%%%%%%%%%%%%%%%%%%%%%%%%%%%%%%%

The strong impact of the Fermi surface on the conductivity anisotropy in PtCoO$_2$ is explained by
the band resolved in-plane and out-of-plane components of the electrical
conductivity tensor of PtCoO$_2$ (Fig.\ \ref{fig:sigbands}). We see that for tensile strains
only the 16th band contributes since it is still the only one crossing the Fermi surface.
Its in-plane component remains essentially constant, while the out-of-plane component is
responsible for the exponential decrease of the conductivity anisotropy shown
in Fig.\ \ref{fig:Sigma}.
The additional bands start
contributing to the conductivity tensor below an in-plane lattice constant
$a$$\,<\,$$2.80\,$\AA{} (15th band) and
$a$$\,<\,$$2.83\,$\AA{} (17th band), respectively, see Fig.\ \ref{fig:sigbands}.
For tensile strains, the conductivity is maintained by the 16th band, alone.
Its strong decrease can be related to an increasingly better definition of the hexagonal
corners of the Fermi-surface along the $k_z$, as visible from the comparison of the Fermi surface
under tensile strain with the equilibrium Fermi surface (Fig.\ \ref{fig:FSstrain}a-\ref{fig:FSstrain}c, center row).
This corresponds to a diminished band velocity
component in $z$-direction, which consequently reduces the respective element of the conductivity tensor.
In a simple geometric picture, this complies with the decreasing width of the conducting PtO$_2$-layer,
while the separation between O and Co and thus the width of the insulating layer becomes larger.

Recent experiments obtained for the anisotropy $\sigma_{\rm xx}$/$\sigma_{\rm zz}$ of PdCoO$_2$ values
ranging from 150 (Ref.\ \onlinecite{cn:Takatsu07JPSJ}), 280 (Ref.\ \onlinecite{cn:Daou15PRB}) to
400 (Ref.\ \onlinecite{cn:Hicks12PRL}).
This large discrepancy has been noticed earlier and related to the treatment
of umklapp processes and the anisotropy of defect scattering cross sections.\cite{cn:Hicks12PRL}
In particular, the latter aspects point out possible shortcomings of the single relaxation time approximation and
semi-classical Boltzmann transport theory.
Computing the anisotropy ratio from Fig.\ \ref{fig:Sigma} assuming a single constant $\tau$,
the in-plane conductivity is a factor 30 larger than
out-of-plane.
This ratio can be somewhat increased by applying static Coulomb correlations within the GGA+$U$ scheme
(see the supporting information\cite{cn:supplement} for more details) but it never becomes comparable to the
experimental results. We take this as another indication that  a directional dependence of $\tau$, which has been
neglected so far, should enter the anisotropy ratio.

The diagonal structure of the transport tensors allows us to pragmatically circumvent this problem
by introducing two effective, temperature dependent relaxation time constants
for in-plane and out-of-plane processes, $\tau_{\rm xx}$ and
$\tau_{\rm zz}$, respectively, which we obtain by comparison with the experimental conductivities.
This effectively re-introduces the $k$-vector dependence of $\tau$, which is discarded in Boltzmann theory
within the common single relaxation time approximation.
As for the anisotropy, experiment offers a considerable span of results also for the in-plane experimental
conductivities $\sigma_{\rm xx}$, ranging from $14.5\,\mu\Omega^{-1}$m$^{-1}$ (Ref.\ \onlinecite{cn:Takatsu07JPSJ}),
over $32.3\,\mu\Omega^{-1}$m$^{-1}$ (Ref.\ \onlinecite{cn:Daou15PRB}) and
$38.5\,\mu\Omega^{-1}$m$^{-1}$ (Ref.\ \onlinecite{cn:Hicks12PRL}) to $50\,\mu\Omega^{-1}$m$^{-1}$ (Ref.\ \onlinecite{cn:Rogers71}).
This results in relaxation time constants $\tau_{\rm xx}(300\,{\rm K})$ of $46$\,fs, $101$\,fs,
$121$\,fs and $157$\,fs, respectively.
These comparatively large values are consistent with the previous assessment of Ong {\em et al.},\cite{cn:Ong10PRL}
obtained in a similar fashion by comparing Boltzmann theory and experimental conductivity data.
From the residual in-plane resistivities for $T\to 0$, Hicks {\em et al.}\cite{cn:Hicks12PRL} concluded on an extremely
large transport mean free path $l_{\rm MFP}$$\,=\,$$20\,\mu$m. This is considerably larger than the earlier estimate
$l_{\rm MFP}$$\,=\,$$60\,$\AA{} of Noh {\em et al.}\cite{cn:Noh09PRL} based on the peak width obtained from
ARPES. These measurements also yield a value for the in-plane carrier velocity of $4.96$\,eV\,\AA{}\,$\hbar^{-1}$
which is consistent with first-principles results.\cite{cn:Ong10PRL}
Thus, relaxation time constants obtained from experiment were either significantly smaller, such as
the $7.6$\,fs estimated by Noh {\em et al.}\cite{cn:Noh09PRL} 
or larger, based on the de-Haas-van-Alphen measurements of Hicks {\em et al.}\cite{cn:Hicks12PRL} or the analysis of
anomalous magnetoresistance.\cite{cn:Takatsu13PRL}
For the out of plane direction, ambient $\sigma_{\rm zz}$ was measured as
$0.096\,\mu\Omega^{-1}$m$^{-1}$ (Ref.\ \onlinecite{cn:Hicks12PRL}),
$0.097\,\mu\Omega^{-1}$m$^{-1}$ (Ref.\ \onlinecite{cn:Takatsu07JPSJ}) and
$0.115\,\mu\Omega^{-1}$m$^{-1}$ (Ref.\ \onlinecite{cn:Daou15PRB}), corresponding to
$\tau_{\rm zz}(300\,{\rm K})$$\,=\,$$9$\,fs and $11$\,fs, respectively, which yields a more uniform picture
than the in-plane-case.

Most experiments were carried out at room temperature and below, only Takatsu {\em et al.} performed conductivity
measurements up to $500$\,K. Thus, only Ref.\ \onlinecite{cn:Takatsu07JPSJ} offers a reasonable possibility
to extrapolate the experimental conductivities to $T$$\,=\,$$600$\,K.
We find then $\sigma_{\rm xx}$$\,=\,$$5\,\mu\Omega^{-1}$m$^{-1}$
and  $\sigma_{\rm zz}$$\,=\,$$0.04\,\mu\Omega^{-1}$m$^{-1}$, which corresponds to
$\tau_{\rm xx}(600\,{\rm K})$$\,=\,$$16$\,fs and $\tau_{\rm zz}(600\,{\rm K})$$\,=\,$$4$\,fs.
For PtCoO$_2$, Rogers {\em et al.}\cite{cn:Rogers71}
reported conductivities $\sigma_{\rm zz}$$\,=\,$$33\,\mu\Omega^{-1}$m$^{-1}$
and $\sigma_{\rm zz}$$\,=\,$$0.1\mu\Omega^{-1}$m$^{-1}$.
This leads to $\tau_{\rm xx}(300\,{\rm K})$$\,=\,$$70$\,fs and
$\tau_{\rm zz}(300\,{\rm K})$$\,=\,$$12$\,fs,
which is fairly close to the relaxation times obtained for PdCoO$_2$.

\subsection{Thermoelectric performance}\label{sec:trans}

%%%%%%%%%%%%%%%%%%%%%%%%%%%%%%%%%%%%%%%%%%%%%%%%%%%%%%%%%%%%%%%%%%%%%
\begin{figure} % Fig. 7
  \begin{center}  
\includegraphics[width=0.8\columnwidth]{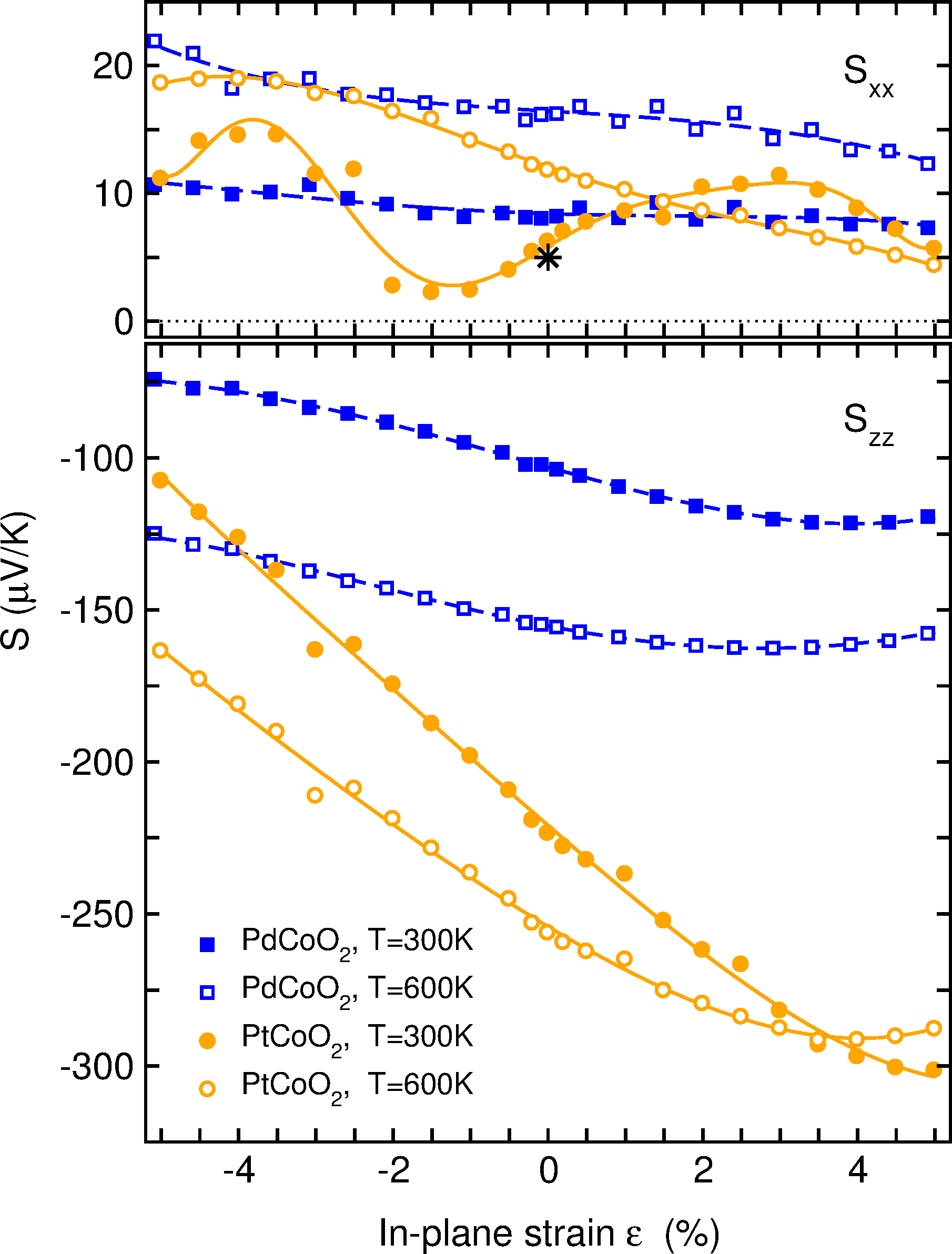} %{fig/PdPt-S-Strain.png}
  \end{center}
  \caption{(color online)
    Diagonal elements $S_{\rm xx}$ (upper panel) and $S_{\rm zz}$ (lower panel) of the
    tensorial thermopower of PdCoO$_2$ (blue dashed lines and squares) and PtCoO$_2$
    (orange solid lines and circles) as a function of the epitaxial in-plane strain for two
    different temperatures $T$$\,=\,$$300\,$K (open symbols) and
    $T$$\,=\,$$600\,$K (filled symbols). The black star in the upper panel marks the experimental in-plane
    thermopower of PdCoO$_2$ at $T$$\,=\,$$300\,$K taken from Ref.\ \protect\onlinecite{cn:Daou15PRB}.}
\label{fig:Seebeck}
\end{figure}
%%%%%%%%%%%%%%%%%%%%%%%%%%%%%%%%%%%%%%%%%%%%%%%%%%%%%%%%%%%%%%%%%%%%%

The thermopower of PdCoO$_2$ was first reported
by Yagi {\em et al.}\cite{cn:Yagi00KEM} for temperatures above $500$\,K.
Since the authors measured on polycristalline materials, the highly anisotropic
behavior of this quantity was overseen. Later, Hasegawa {\em et al.}\cite{cn:Hasegawa02SSC}
measured the thermopower at lower temperatures, again on powdered and sintered crystals. The authors obtained a positive
value of $S$$\,=\,$$2\ldots 4\,\mu$V\,K$^{-1}$ at room temperature, which is comparable to conventional metals.
In a very recent study the in-plane thermopower $S_{\rm xx}$ and thermal conductivity
of PdCoO$_2$ was measured for temperatures below $300$\,K.\cite{cn:Daou15PRB}
$S_{\rm xx}$ exhibits a change of sign below $T$$\,<\,$$100$\,K,
but approaches the expected linear temperature dependence at higher
temperatures, reaching $S_{\rm xx}$$\,=\,$$5\,\mu$V\,K$^{-1}$ at ambient conditions.

Fig.\ \ref{fig:Seebeck} shows the calculated thermopower (Seebeck coefficient) $S$ as a function of the
epitaxial strain for both directions and two different temperatures $T$$\,=\,$$300\,$K and
$T$$\,=\,$$600\,$K.
For both systems, we confirm the marked anisotropy in the Seebeck coefficient,
which is significantly more pronounced for
PtCoO$_2$ than for PdCoO$_2$ and increases with temperature.
It is positive for the in-plane component $S_{\rm xx}$, which indicates a higher mobility of p-type carriers (holes),
while large negative values for  $S_{\rm xx}$ point our the dominance of n-type carriers (electrons)
for transport in out-of-plane direction.
The estimate for $S_{\rm xx}$ of unstrained PdCoO$_2$ from Boltzmann theory at room temperature
is approximately 1.6 times larger than the experimental value published in Ref.\ \onlinecite{cn:Daou15PRB}.
This might still be considered a reasonable agreement keeping in mind the simplifications of Boltzmann transport
theory and experimental difficulties in obtaining this quantity.
Previous calculations by Ong {\em et al.}\cite{cn:Ong10PRL,cn:Ong10PRB} report moderate positive
values in-plane for $S_{\rm xx}$, while the out of plane component $S_{\rm zz}$ provides a
large negative contribution.
Our values compare well to the results of Ong {\em et al.},\cite{cn:Ong10PRL} who carried out
their calculations with the Wien2k code for the experimental lattice constant, which
differs slightly from our setup.

We now turn to the effect of strain, which is varied from -5\% compressive to +5\% tensile strain.
The  in-plane thermopower $S_{\rm xx}$ and its variation under strain remains moderate in absolute numbers.
We achieve changes in the range of $10$$\ldots$$12$$\,\mu$V/K at $T$$\,=\,$$600\,$K as compared to
$\varepsilon$$\,=\,$$0$
for both systems. Nevertheless, as the absolute values are small, maximum strain corresponds to a relative change
by a factor of two for the Pt-based oxide. The oscillations observed at lower temperatures
are a consequence of the electronic topological transition
related to the two additional bands consecutively crossing the Fermi level.
For the out-of-plane component $S_{\rm zz}$, we find much larger changes in absolute numbers, in particular, around room temperature.
Here, the tensile strain yields a relative increase of $60$\,\% to $180$\,\%, which corresponds
to rather significant changes in absolute numbers
of $-47\,\mu$V/K and $-196\,\mu$V/K  for PdCoO$_2$ than for PtCoO$_2$, respectively.

In particular, for PtCoO$_2$, the variation of the out-of-plane thermopower, $S_{\rm zz}$, with strain bears
close similarities with the
logarithm of the strain dependence of the conductivity element $\sigma_{\rm zz}$.
Such a relation is motivated by the textbook formula of Mott,
which connects the scalar thermopower $S$
of an isotropic system with the logarithmic derivative of the (scalar)
conductivity $\sigma$ with respect to the chemical potential $\mu$:
\begin{equation}
  S =
\frac{\pi^2k_{\rm B}^2T}{3e}\,\frac{\partial}{\partial\mu}
\ln\left[\sigma(\mu)\right]\,,
\end{equation}
We therefore conclude that the large strain variations in the thermopower
originate from the corresponding changes in the conductivity discussed above.
The negative sign of $S_{\rm zz}$ is related to the decrease of $\sigma_{\rm zz}$
with increasing chemical potential $\mu$ at a given strain,\cite{cn:Kim14JPSJ}
while the strong variation of the magnitude of $|S_{\rm zz}|$ with strain relates inversely
to the change in conductivity.

%%%%%%%%%%%%%%%%%%%%%%%%%%%%%%%%%%%%%%%%%%%%%%%%%%%%%%%%%%%%%%%%%%%%%
\begin{figure} % Fig. 8
  \begin{center}  
\includegraphics[width=0.85\columnwidth]{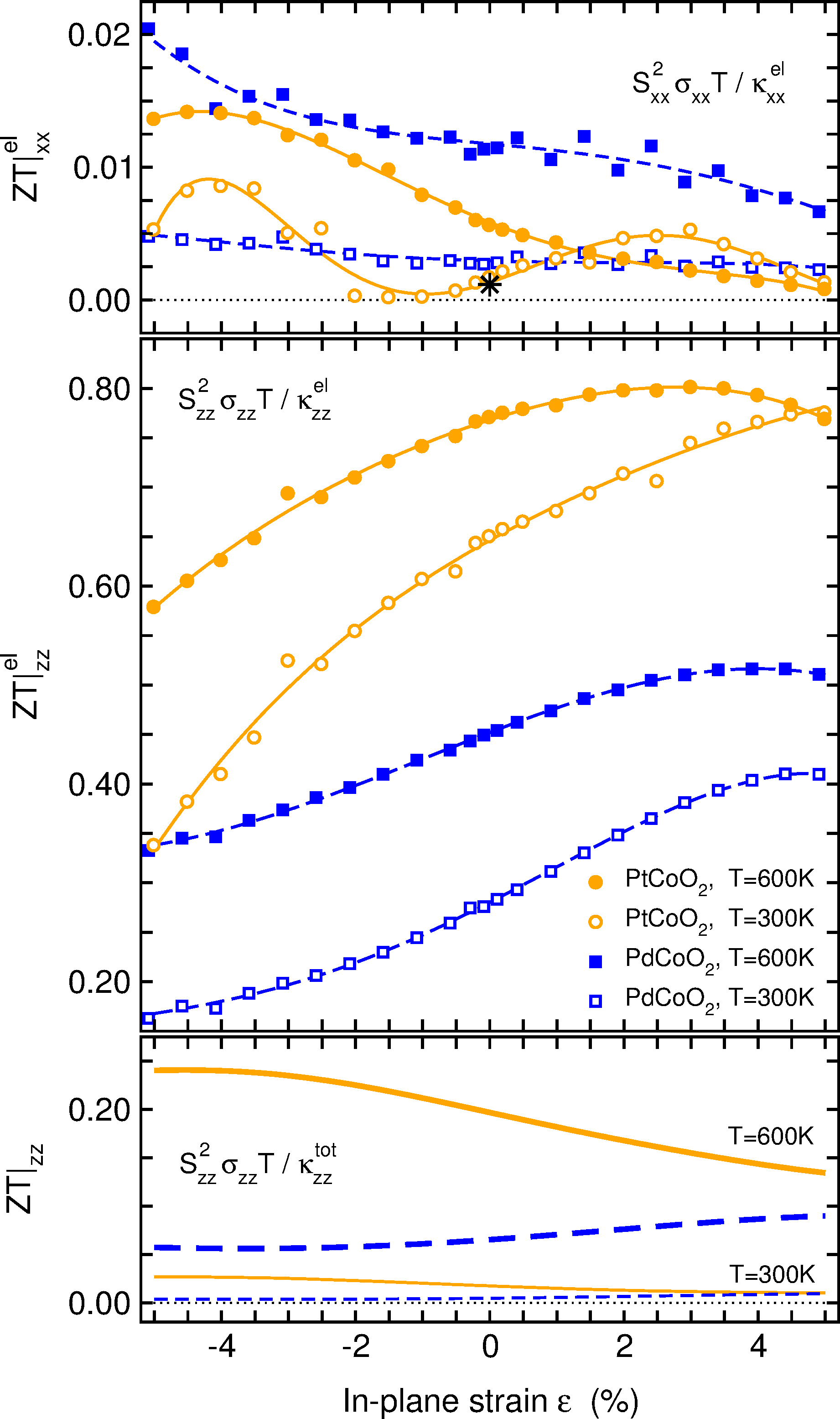} %{fig/PdPt-ZTall-Strain.png}
  \end{center}
  \caption{(color online)
    Thermoelectric figure of merit, $ZT$,
    calculated from the diagonal elements of the transport tensors
    (upper panel, xx elements; lower panel zz elements, symbols and colors as in the preceding figures).
    Thin lines and data points in the upper and midel panel denote an upper boundary  $ZT|^{\rm el}$ as
    lattice thermal conductivity is neglected. The bottom panel shows an estimate of the total $ZT$
    using the experimental $\kappa_{zz}^{\rm ph}$ for PdCoO$_2$ from
    Ref.\ \onlinecite{cn:Daou15PRB} (see text)
    and the relaxation time constant $\tau$$\,=\,$$10$\,fs (same values for both compounds and all strains).
    The thin lines refer to $T$$\,=\,$300\,K and the thick lines to $T$$\,=\,$600\,K, while broken lines denote PdCoO$_2$
    and solid lines PtCoO$_2$.
    The black star in the upper panel marks the experimental in-plane $ZT$ of PdCoO$_2$ at $T$$\,=\,$$300\,$K
    calculated from the experimental data of Daou and coworkers.\protect\cite{cn:Daou15PRB}}
\label{fig:ZT}
\end{figure}
%%%%%%%%%%%%%%%%%%%%%%%%%%%%%%%%%%%%%%%%%%%%%%%%%%%%%%%%%%%%%%%%%%%%%

The figure of merit of a material with respect to its thermoelectric performance
is given by the dimension-less number
$ZT$$\,=\,$$S^2\sigma\,T/(\kappa^{\rm el}+\kappa^{\rm ph})$,
where $\kappa^{\rm el}$ and
$\kappa^{\rm ph}$ are the electronic and lattice thermal conductivity,
respectively.
Apart from $\kappa^{\rm ph}$, all quantities are accessible
within our approach.
However, $\kappa^{\rm ph}$ becomes the dominant contribution,
when the electrical conductivity is minute, e.g. in semiconductors or insulators,
and will thus be relevant for the out-of-plane transport.
The calculation of $\kappa^{\rm ph}$ from first-principles
requires the determination of the anharmonic contributions to lattice dynamics,
which is beyond the scope of the present work.
However, the total and lattice contributions to the thermal conductivity of PdCoO$_2$ were recently obtained experimentally
by Daou {\em et al.}\cite{cn:Daou15PRB} for temperatures up to $320$\,K.
For the in-plane case, the authors reported a rather large
total thermal conductivity $\kappa^{\rm tot}_{xx}$ of
$250$\,W\,K$^{-1}$m$^{-1}$, while the total out-of-plane conductivity $\kappa^{\rm tot}_{zz}$ amounts to approximately
$70$\,W\,K$^{-1}$m$^{-1}$ at ambient conditions. According to the Wiedemann-Franz law, the electronic contribution
$\kappa^{\rm el}_{zz}$ is expected to be at least one order of magnitude smaller, based on the very low out-of-plane conductivity.
This yields thus a direct estimate of the lattice thermal conductivity $\kappa^{\rm ph}_{zz}$.

According to the Wiedemann-Franz law, the electronic conductivities $\kappa^{\rm el}$
increase with temperature, whereas for the lattice thermal conductivity $\kappa^{\rm ph}$ a fast decrease is
expected according to the Debye-Callaway model.
For the in-plane component of $ZT$, we can neglect  $\kappa^{\rm ph}$
as a first approximation since it is significantly exceeded by $\kappa^{\rm el}$.
The corresponding quantity, which we denote by $ZT|^{\rm el}$, provides thus an upper limit
for the true $ZT$.  According to the large thermal conductivity, the in-plane values turn out to be
prohibitively low from the application point of view.
Since the strain dependence of $\sigma$ and $\kappa^{\rm el}$ essentially
cancels out, the strain dependence of $ZT|_{\rm xx}^{\rm el}$ is dominated by the contribution from
$S_{\rm xx}^2$ (cf.\ Fig.\ \ref{fig:ZT}).
This is also the case for $ZT|_{\rm zz}^{\rm el}$. However, since the thermal conductivity is
dominated by the lattice contribution, we cannot use $ZT|_{\rm zz}^{\rm el}$ to assess the thermoelectric
performance of the material. Instead, we determine the complete $ZT|_{\rm zz}$
with the help the experimental value $\kappa_{\rm zz}^{\rm ph}$ of Ref.\ \onlinecite{cn:Daou15PRB}
for $T$$\,=\,$$300$\,K and  $\kappa_{\rm zz}^{\rm ph}$$\,=\,$$10$\,W\,K$^{-1}$m$^{-1}$ for $T$$\,=\,$$600$\,K,
which can be estimated from the fit to the Debye-Callaway model provided in Ref.\ \onlinecite{cn:Daou15PRB}.
In addition, we use an average $\tau_{\rm zz}$$\,=\,$$10$\,fs for $T$$\,=\,$$300$\,K, while
for $T$$\,=\,$$600$\,K,
we take $\tau_{\rm zz}$$\,=\,$$4$\,fs, as discussed in Sec.\ \ref{sec:conduct}.
As there is only sufficent data for PdCoO$_2$, we use the same parameters for both systems.
The strain dependece of $ZT|_{\rm zz}$ is plotted as thick dashed (PdCoO$_2$) and solid
(PtCoO$_2$) lines in the bottom part of Fig.\ \ref{fig:ZT}.
Since the dominant contribution to thermal conductivity comes from the lattice, which we assume constant
for all $\varepsilon$, the strain dependence of $ZT|_{\rm zz}$ rather resembles the power factor (see supplementary
information\cite{cn:supplement}).
As a general trend, we see that $ZT|_{\rm zz}$ increases strongly with temperature,
while PtCoO$_2$ clearly exceeds the performance of PdCoO$_2$. Thus, only PtCoO$_2$ can reach a reasonable figure of merit
of $ZT|_{\rm zz}$$\,=\,$$0.25$ at compressive epitaxial strains and sufficiently high temperatures, which is a consequence of
the electronic topological transition.

\section{Conclusions}\label{sec:conclusions}

Based on first-principles calculations in combination with Boltzmann transport theory in the
single relaxation time approximation, we provide a systematic analysis
of the electronic structure and anisotropic transport properties of the delafossites
PdCoO$_2$ and PtCoO$_2$ under epitaxial strain.
We demonstrate that despite the large similarities in both systems concerning their structural properties
that PtCoO$_2$ has -- unlike PdCoO$_2$ --
the propensity to undergo an electronic topological transition, which might be
triggered by a realistic compressive epitaxial strain. In turn, by expanding the in-plane lattice constant, it exhibits
a dimensional crossover from a threedimensional open Fermi surface, which touches the zone boundary in-plane and out of plane,
to a nearly perfect twodimensional electronic system, with a closed hexagonal Fermi-surface with
perfectly flat sides extending in $k_z$-direction. 

Comparing the in-plane and out-of-plane conductivies from our
calculations based on GGA correlation and the GGA+$U$ approach with experiment, 
we propose that two different relaxation times %, which account for the relevant scattering mechanisms,
must be introduced to describe in-plane and out-of-plane transport appropriately in Boltzmann transport theory.
We consider this approach justified in the present case due to the essentially
diagonal structure of the transport tensors.

We finally predict that, despite the apparent similarities of both oxides,
PtCoO$_2$ exhibits a much better thermoelectric performance
than PdCoO$_2$ and might thus be a better model system for applications.
Our analysis includes the thermoelectric figure of merit $ZT$,
which we obtain by combining theoretical results with recently available experimental data
for the lattice thermal conductivity of PdCoO$_2$.\cite{cn:Daou15PRB}
The presence of the topological transition significantly helps in improving this number,
since it increases the out-of-plane conductivity, which can be tuned efficiently by
an external control parameter, such as epitaxial strain.
In this way, we can balance the contributions of electronic and lattice thermal conductivity effectively, while
still allowing for sufficiently large absolute values of the thermopower.
With this strategy we arrive at a reasonbale figure of merit
of $ZT$$\,=\,$$0.25$ at $T$$\,=\,$$600$\,K in out-of-plane direction for a system under compressive in-plane strain.
For tensile strains the out-of-plane conductivity becomes very low and $ZT$ small,
since the lattice thermal conductivity dominates. But since the thermopower is almost three
times larger than for compressive strains, appropriate doping of carriers may still
improve the thermoelectric performance reasonably.

In conclusion, the results for PtCoO$_2$ as a model system confirm that metallic materials
characterized by a quasi-twodimensional Fermi surface may be well suited for thermolectric applications.
However, we emphasize that for a reasonable performance
the two-dimensional shape must not be too perfect.
In the present case, epitaxial strain was used effectively as an external parameter 
to obtain a certain level of imperfection that optimizes the relation of
electrical and thermal conductivities in the figure of merit.

%%%%%%%%%%%%%%%%%%%%%%%%%%
\section*{Acknowledgments}
%%%%%%%%%%%%%%%%%%%%%%%%%%
The authors gratefully acknowledge
useful discussions with H.-U.\ Habermeier, B.\ Keimer (Stuttgart) and M.\ Verstraete (Li\`{e}ge).
Financial support was granted
by the Deutsche Forschungsgemeinschaft in the framework of
SFB/TRR 80 (project G8).

%\bibliographystyle{apsrev}
%\bibliography{DFT,own,PdCoO2,TE}

%%%%%%%%%%%%%%
\end{document}